\documentclass[10pt, twocolumn]{article}

\usepackage[T1]{fontenc}
\usepackage[utf8]{inputenc}
\usepackage{mathptmx}
\usepackage[expansion=false]{microtype}

\usepackage{amsmath,amssymb}

\usepackage[top=0.85in, bottom=0.85in, left=0.75in, right=0.75in]{geometry}
\usepackage{setspace}
\usepackage{parskip}
\usepackage{titlesec}
\titleformat{\section}{\normalsize\bfseries}{\thesection.}{0.5em}{}
\titleformat{\subsection}{\small\bfseries}{\thesubsection}{0.5em}{}
\titlespacing*{\section}{0pt}{6pt}{3pt}
\titlespacing*{\subsection}{0pt}{5pt}{2pt}

\usepackage{abstract}

\usepackage{booktabs}
\usepackage{array}
\usepackage{tabularx}
\usepackage{makecell}

\usepackage{graphicx}
\usepackage{float}
\usepackage[font=small, labelfont=bf, skip=4pt]{caption}
\usepackage{tikz}
\usetikzlibrary{
  arrows.meta,
  shapes.geometric,
  shapes.misc,
  positioning,
  fit,
  calc,
  backgrounds,
  decorations.pathreplacing,
  decorations.markings
}

\usepackage{hyperref}
\hypersetup{colorlinks=true, linkcolor=blue!60!black,
            urlcolor=blue!60!black, citecolor=black}

\usepackage{enumitem}
\setlist[enumerate]{itemsep=1pt, topsep=2pt, parsep=0pt}

\usepackage{xcolor}
\usepackage[normalem]{ulem}

\tikzset{
  sbox/.style={draw, rectangle, rounded corners=2pt,
               minimum width=#1, minimum height=0.55cm,
               align=center, font=\tiny, inner sep=2pt},
  sbox/.default=1.1cm,
  arr/.style={->, >=Stealth, thick},
  darr/.style={->, >=Stealth, dashed},
}

\title{\textbf{Write-Read Decoupling in Modern Large-Scale Search Engines:
       Architectures, Techniques, and Emerging Approaches}}
\author{%
  Xin Liang\textsuperscript{*} \quad
  Qing Yang \quad
  Wenru Qiu \quad
  Wenjie Mao \quad
  Tianyu Ma \quad
  Minghui Zhu \quad
  Nan Wang \\[4pt]
  \small \textsuperscript{*}\texttt{xinlmain@gmail.com}
}
\date{}

\begin{document}

\twocolumn[
  \maketitle
  \vspace{-18pt}
  \begin{abstract}
    \noindent
      Large-scale search engines face a fundamental tension: the index must be
      updated frequently to maintain freshness, yet updates create resource
      contention that inflates query latency.  In the dominant Lucene-based
      architecture, segment merges triggered by writes compete with concurrent
      queries for CPU cycles, disk I/O bandwidth, and operating-system page
      cache---a problem we term \emph{write-read contention}.  This survey
      systematically examines the architectural solutions that industry and
      academia have developed to decouple write pressure from read latency.  We
      identify five principal patterns: (i)~node-level read-write separation;
      (ii)~compute-storage separation; (iii)~full in-memory indexing;
      (iv)~log-structured write paths; and (v)~in-place partial updates.  We
      survey representative systems including Elasticsearch, LinkedIn Galene,
      Uber Sia, Quickwit, Alibaba Havenask, Algolia, Milvus, and Vespa, and
      discuss an emerging synthesis---the ScaleSearch architecture---that
      combines compute-storage separation with full in-memory indexing and
      dedicated write nodes.  A key contribution of ScaleSearch is
      \emph{per-field update routing}: each field is assigned its own Kafka
      topic and update path, allowing scalar fields (price, stock, tags) to be
      updated in-place in $O(1)$ RAM with immediate visibility while full-text
      fields follow the segment-based compute-storage path.  We conclude with
      open challenges in hybrid vector-and-full-text retrieval, serverless
      deployments, and AI-integrated search.

  \end{abstract}
  \vspace{10pt}
  \hrule
  \vspace{8pt}
]

\section{Introduction}

Search engines are foundational infrastructure for consumer-facing applications.
In e-commerce, advertising, recommendation, and social discovery, the search
engine must simultaneously satisfy two stringent requirements: (1)~query latency
well under 100\,ms, and (2)~index freshness at the minute or sub-minute level.
Satisfying both requirements concurrently is a deep engineering challenge.

The difficulty arises from a structural conflict in the dominant indexing
technology.  Apache Lucene~\cite{bib:lucene4} organizes the inverted index as a
collection of immutable \emph{segments}.  New documents are flushed to a new
on-disk segment (a \emph{refresh}), making them searchable; over time, small
segments are merged into larger ones (a \emph{segment merge}).  Both operations
are I/O- and CPU-intensive.  When executed concurrently with search queries on
the same node, they compete for the same hardware resources, producing elevated
and unpredictable query latency.

This tension is structural.  The Lucene segment model is an instance of the
Log-Structured Merge-tree (LSM-tree)~\cite{bib:lsm}, and the write-read
trade-off is a fundamental property of LSM-based storage.  The write
amplification factor (WAF)---the ratio of bytes actually written to disk per
byte of user data---is typically $10\text{--}30\times$ for Elasticsearch under
continuous ingestion~\cite{bib:qader2018}.  In Elasticsearch's
\emph{shared-nothing} architecture, scaling out does not help: every replica
independently re-runs tokenization, segment construction, and merges, multiplying
total write work by $(1 + \text{replicas})$~\cite{bib:esnodes}.

The past decade has seen a burst of architectural innovation.  Cloud
infrastructure has enabled new approaches: cheap durable object storage provides
a shared medium that decouples indexing from serving; large-memory servers make
full in-memory indexes economically viable; and container orchestration allows
read and write workloads to scale independently.

This survey organizes and analyzes these innovations.  Our contributions are:
(1)~a precise characterization of the contention problem;
(2)~a taxonomy of five decoupling patterns with representative systems;
(3)~a comparative analysis of freshness-latency-cost trade-offs;
(4)~a description of the ScaleSearch architecture as a synthesis, with
particular emphasis on its \emph{per-field update routing} mechanism that
assigns each field an independent update path matched to its access pattern;
and (5)~an outlook on vector search, serverless, and AI-augmented retrieval.

\section{Background}

\subsection{The Inverted Index}

An inverted index maps each term $t$ to a \emph{postings list}: an ordered
sequence of document identifiers (docIDs) that contain $t$, augmented with term
frequencies and positional data.  Building a static index is well understood:
Single-Pass In-Memory Indexing (SPIMI)~\cite{bib:manning2008} accumulates
postings in memory and writes sorted inverted lists to disk in one pass.  The
challenge is \emph{dynamic} indexing---maintaining the index under continuous
writes while serving concurrent queries.

\subsection{Dynamic Indexing and Logarithmic Merging}

The naive \emph{Immediate Merge} strategy reconstructs the entire index on every
flush, incurring $O(N^2)$ total indexing cost over $N$ documents.  The opposite
extreme, \emph{No-Merge}, bounds indexing cost to $O(N)$ but causes query
overhead to grow linearly with file count.

The practical solution is \emph{logarithmic merging}~\cite{bib:cutting1990,
bib:lester2005}, in which sub-indexes are maintained at exponentially increasing
sizes ($2^0 M$, $2^1 M$, $2^2 M$, $\ldots$); whenever two sub-indexes reach the
same size, they are merged.  This caps total indexing cost at $O(N \log N)$ and
the number of active sub-indexes at $O(\log N)$.  B\"{u}ttcher and
Clarke~\cite{bib:buttcher2005} provide a rigorous characterization of all three
strategies and show that the optimal choice depends on the workload's
query-to-update ratio.

\subsection{The Lucene Segment Model}

Lucene~\cite{bib:lucene4} implements logarithmic merging as its \emph{segment
model}.  The lifecycle is:
\begin{enumerate}
  \item Documents accumulate in a RAM buffer.
  \item \textbf{Refresh}: the buffer is flushed to a new segment in the OS
        page cache, making it searchable without an \texttt{fsync}.
  \item \textbf{Flush}: Lucene performs an \texttt{fsync}, clearing the
        translog.
  \item \textbf{Merge}: background threads coalesce small segments via
        \texttt{TieredMergePolicy}.
\end{enumerate}
Near Real-Time (NRT) search~\cite{bib:nrt} uses
\texttt{DirectoryReader.open(IndexWriter)} to make a refreshed segment
searchable without a full commit.

\subsection{Write Amplification}

Under Elasticsearch's default \texttt{TieredMergePolicy}, empirical measurements
by Qader et al.~\cite{bib:qader2018} show WAF between $10\times$ and $30\times$
for continuous high-rate ingestion.  A system ingesting 10\,MB/s of user data
can generate up to 300\,MB/s of disk write traffic---easily saturating NVMe
drives shared with concurrent queries.

\section{The Write-Read Contention Problem}

\subsection{Mechanisms of Interference}

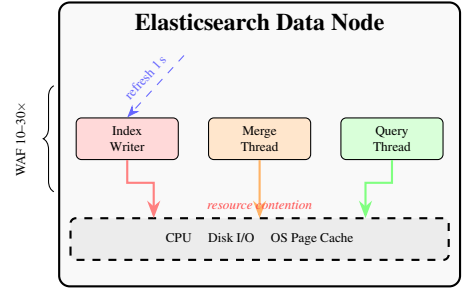
\begin{figure}[t]
\centering
\begin{tikzpicture}
  \draw[thick, rounded corners=4pt, fill=gray!5]
    (-0.15,-0.15) rectangle (5.15,3.6);
  \node[font=\small\bfseries] at (2.5,3.35) {Elasticsearch Data Node};

  \node[sbox=1.35cm, fill=red!15]    (iw) at (0.75,1.8) {Index\\Writer};
  \node[sbox=1.35cm, fill=orange!20] (mt) at (2.5, 1.8) {Merge\\Thread};
  \node[sbox=1.35cm, fill=green!15]  (qt) at (4.25,1.8) {Query\\Thread};

  \draw[thick, dashed, fill=gray!15, rounded corners=2pt]
    (0.0,0.2) rectangle (5.0,0.75);
  \node[font=\tiny] at (2.5,0.48) {CPU \quad Disk I/O \quad OS Page Cache};

  \draw[arr, red!50]    (iw.south) -- ++(0,-0.3) -| (1.1,0.75);
  \draw[arr, orange!60] (mt.south) -- (2.5,0.75);
  \draw[arr, green!50]  (qt.south) -- ++(0,-0.3) -| (3.9,0.75);

  \node[font=\tiny, red!70, above] at (2.5,0.75)
    {\textit{resource contention}};

  \draw[decorate, decoration={brace, amplitude=4pt, raise=2pt}]
    (-0.15,1.1) -- (-0.15,2.5);
  \node[font=\tiny, rotate=90] at (-0.65,1.8) {WAF $10$--$30\times$};

  \draw[darr, blue!60] (1.5,2.9) -- node[font=\tiny,above,sloped]{refresh 1\,s}
    (iw.north);
\end{tikzpicture}
\caption{Write-read contention on a shared Elasticsearch node.  Index
  writes, background segment merges, and queries compete for the same CPU,
  disk I/O, and OS page cache.}
\label{fig:contention}
\end{figure}

Figure~\ref{fig:contention} illustrates the three main interference pathways
on a shared Elasticsearch node.

\textbf{CPU contention.}  Segment merging is CPU-intensive: it re-encodes
posting lists, sorts docID arrays, and rebuilds FST-based term dictionaries.
Merge threads compete with query-processing threads for CPU time, inflating
tail latency.

\textbf{I/O contention and cache eviction.}  Merges generate large sequential
I/O that displaces \emph{hot} search segments from the OS page cache---the
primary read cache for Lucene's \texttt{MMapDirectory}.  Post-merge, the first
queries against the new segment must re-read data from disk, causing latency
spikes documented by McCandless~\cite{bib:mccandless2011}.

\textbf{Segment explosion.}  With the default 1-second refresh interval,
Elasticsearch creates one new segment per second per active shard.  Hundreds of
small segments accumulate before merges coalesce them.  Each query scans all
active segments; the resulting heap pressure from per-segment metadata and the
cost of scanning many short postings lists degrades throughput by
$2\text{--}5\times$ compared to a fully merged index~\cite{bib:eshards}.

\subsection{Shared-Nothing Amplification}

In Elasticsearch's shared-nothing model, every data node independently maintains
its assigned shards on local disk and independently re-runs the full indexing
pipeline on every replica---multiplying total write work by
$(1 + \text{replicas})$~\cite{bib:esnodes}.  Scaling out does not reduce write
pressure per shard; newly added nodes receive reassigned primaries and must
immediately absorb a full write load.  The ratio of write overhead to query
capacity thus remains approximately constant as the cluster grows.

\subsection{The Freshness-Latency-Cost Triangle}

The three objectives---low query latency, high update freshness, and low
operational cost---form a triangle where improving any one dimension tends to
worsen the others.  Reducing the refresh interval improves freshness but
increases segment count and degrading latency.  Adding more nodes improves
capacity but raises total I/O through replica write amplification.  No single
configuration of the standard Elasticsearch architecture can simultaneously
optimize all three.  The solutions in Section~\ref{sec:solutions} can be
understood as strategies for escaping this triangle by breaking the coupling
between the indexing and serving code paths.

\section{Architectural Solutions}
\label{sec:solutions}

\subsection{Node-Level Read-Write Separation}

The most direct response is to stop co-locating the two workloads.  Dedicated
nodes handle indexing exclusively; other nodes serve queries exclusively.  The
index propagates from indexing nodes to serving nodes as immutable files.

\textbf{LinkedIn Galene}~\cite{bib:galene} maintains strict separation between
\emph{Indexer} nodes (consuming Kafka streams, building Lucene segments, and
periodically force-merging snapshots) and \emph{Searcher} nodes (loading and
serving snapshots).  Searcher nodes never perform indexing, completely
eliminating merge overhead from the query path.

\textbf{Uber Sia}~\cite{bib:sia} extends this pattern.  Dedicated Ingestion
Services (Apache Flink jobs) build segments offline, apply force-merges, and
publish to object storage (HDFS/GCS).  Stateless Searcher nodes pull pre-built
segments.  Zero live indexing occurs on serving nodes.

Elasticsearch supports a limited form through tiered node roles~\cite{bib:esnodes}:
dedicated ingest nodes pre-process documents; hot-tier nodes host
write-intensive primaries; shard allocation filters pin read replicas to
read-optimized nodes.

\emph{Trade-offs.}  Propagation latency (minutes in snapshot-based systems)
is traded for query latency stability.  Merges still occur on indexing nodes,
just without co-located queries.

\subsection{Compute-Storage Separation}
\label{sec:css}

\begin{figure}[t]
\centering
\begin{tikzpicture}
  \node[sbox=1.0cm, fill=gray!15] (client) at (0,0)   {Client\\Writes};
  \node[sbox=1.1cm, fill=orange!25, right=0.3cm of client] (inode) {Indexing\\Node};
  \node[sbox=1.2cm, fill=blue!20, minimum height=1.3cm,
        right=0.4cm of inode] (s3) {Object\\Storage\\(S3)};
  \node[sbox=1.1cm, fill=green!20, right=0.4cm of s3, yshift= 0.55cm] (sn1) {Search\\Node};
  \node[sbox=1.1cm, fill=green!20, right=0.4cm of s3]                  (sn2) {Search\\Node};
  \node[sbox=1.1cm, fill=green!20, right=0.4cm of s3, yshift=-0.55cm]  (sn3) {Search\\Node};

  \draw[arr] (client) -- (inode);
  \draw[arr, dashed, orange!70] (inode) -- node[font=\tiny,above]{upload} (s3);
  \draw[arr, green!60] (s3) -- node[font=\tiny,above,sloped]{fetch/cache} (sn1);
  \draw[arr, green!60] (s3) -- (sn2);
  \draw[arr, green!60] (s3) -- (sn3);

  \node[font=\tiny, gray, below=1pt of inode] {stateful};
  \node[font=\tiny, gray, below=1pt of sn2]   {stateless};
\end{tikzpicture}
\caption{Compute-storage separation.  Indexing nodes write Lucene
  segments to shared object storage; stateless search nodes fetch and cache
  data on demand.  The two tiers share no hardware resources and scale
  independently.}
\label{fig:css}
\end{figure}
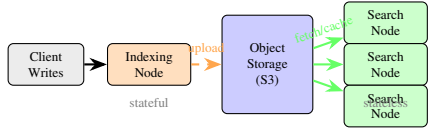

A more radical decoupling eliminates per-node local storage entirely.
The authoritative index resides in shared durable object storage (S3, GCS,
Azure Blob).  Compute nodes are stateless: they cache data locally but hold no
authoritative state, and can be added or removed without data migration.
Figure~\ref{fig:css} illustrates the pattern.

\textbf{Quickwit}~\cite{bib:quickwit} organizes the index as immutable
\emph{splits} (1--15\,GB each) stored natively in S3.  Searcher nodes are
fully stateless: they fetch a split's \emph{hotcache} footer ($\approx$60\,ms)
then issue HTTP range requests for the specific byte ranges needed, requiring at
most three network round trips per split per query.  Vectorized async I/O---with
optimal 8--16\,MiB request sizes~\cite{bib:anyblob}---saturates available
bandwidth.  The main limitation is that splits are immutable: update freshness
is bounded by the split seal interval (minutes to hours).

\textbf{Elasticsearch Stateless / Search AI Lake.}  Elastic's ``Search AI
Lake'' (GA Serverless, 2024) makes object storage the primary tier for all
data~\cite{bib:esstateless}.  Compute nodes use multi-tiered caching (memory +
NVMe).  Psaroudakis et al.~\cite{bib:psaroudakis2024} document a batch-commit
format that reduces cloud I/O by $100\times$ while preserving read-after-write
semantics.  Ingestion spikes no longer degrade query latency.

\textbf{Alibaba OpenStore (AliES)}~\cite{bib:openstore} replaces per-node disks
with a shared OSS storage pool.  Multiple replica shards are backed by a single
physical copy, cutting storage by $\approx$50\%.  A dedicated offline Indexing
Service builds and force-merges segments before committing them to the online
cluster, delivering a reported 70\% improvement in write throughput and 99\%
faster node recovery.

\emph{Trade-offs.}  Perfect horizontal scalability and zero write-read
interference, at the cost of cold-start penalty (first queries before the
cache is warm) and minimum freshness latency bounded by segment seal interval.

\subsection{Full In-Memory Indexing}

When query latency requirements are most stringent---sub-10\,ms for
consumer-facing search, sub-5\,ms for ad bidding---even disk-based caching
introduces unacceptable variance.  The solution is to hold the entire index in
DRAM, eliminating disk I/O from the query path entirely.  The benefit extends
beyond I/O speed: full in-memory storage also eliminates the SERDE overhead of
reading compressed on-disk structures, which alone can consume significant CPU
time even when data is page-cache resident.

\textbf{Algolia}~\cite{bib:algolia_engine} holds all index data as RAM-resident
memory-mapped structures in a C++ engine.  Inverted list intersection uses SIMD
instructions; BM25 scoring is replaced by an integer-based tie-breaking scheme
that avoids floating-point computation.  Indexing and search run as separate OS
processes; the search process is assigned higher CPU scheduling priority.  Index
updates are applied to an incremental in-memory copy and then atomically swapped
in via memory-mapped file replacement.  A \emph{generational} strategy
maintains a large structure (old data) and a small structure (recent updates),
merging them via heap-merge when the small structure reaches a size
threshold~\cite{bib:algolia_arch}.

\textbf{Typesense}~\cite{bib:typesense} uses an Adaptive Radix Trie (ART) as
the primary in-memory index.  Insertions are synchronous: the document is
immediately searchable before the HTTP 201 response returns.  Durability is via
Raft consensus.

\textbf{Ximalaya}~\cite{bib:ximalaya} replaced Elasticsearch for ad recall with
a custom in-memory inverted index using Roaring Bitmaps~\cite{bib:roaring1} for
posting lists and term-level locking for concurrent updates, reducing average
query latency from $\approx$50\,ms to under 5\,ms ($10\times$ improvement).

\textbf{Memory footprint engineering.}  \emph{Roaring Bitmaps}~\cite{bib:roaring1,
bib:roaring2} use a three-container hybrid (Array / Bitmap / Run-Length Encoded),
enabling SIMD-accelerated set operations and serving as the default postings
representation in modern Lucene and Elasticsearch.  \emph{Elias-Fano encoding}
provides optimal compression for sorted docID arrays with direct
rank-and-select support~\cite{bib:pibiri2021}.  \emph{Memory-aware sharding}
splits shards when their footprint exceeds a configurable DRAM threshold,
inspired by HBase region splits~\cite{bib:bigtable}.

\emph{Trade-offs.}  Lowest and most predictable query latency, at the cost of
high DRAM requirements (typically $3\text{--}10\times$ raw data size) and slow
cold-start after node restart.

\subsection{Log-Structured Write Paths}

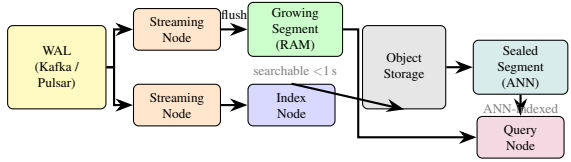
\begin{figure}[t]
\centering
\begin{tikzpicture}[node distance=0.3cm]
  \node[sbox=1.3cm, fill=yellow!30, minimum height=1.1cm] (wal) at (0,0)
    {WAL\\(Kafka /\\Pulsar)};

  \node[sbox=1.15cm, fill=orange!20, right=0.35cm of wal, yshift= 0.5cm] (sn1)
    {Streaming\\Node};
  \node[sbox=1.15cm, fill=orange!20, right=0.35cm of wal, yshift=-0.5cm] (sn2)
    {Streaming\\Node};

  \node[sbox=1.25cm, fill=green!20, right=0.35cm of sn1] (gs)
    {Growing\\Segment\\(RAM)};

  \node[sbox=1.15cm, fill=blue!15, right=0.35cm of sn2] (in1)
    {Index\\Node};

  \node[sbox=1.1cm, fill=gray!20, minimum height=1.1cm,
        right=0.35cm of in1, yshift=0.5cm] (obj)
    {Object\\Storage};

  \node[sbox=1.25cm, fill=teal!15, right=0.35cm of obj, yshift=-0.0cm] (ss)
    {Sealed\\Segment\\(ANN)};

  \node[sbox=1.15cm, fill=purple!15, below=0.28cm of ss] (qn)
    {Query\\Node};

  \draw[arr] (wal.east) -- ++(0.1,0) |- (sn1.west);
  \draw[arr] (wal.east) -- ++(0.1,0) |- (sn2.west);
  \draw[arr] (sn1) -- node[font=\tiny,above]{flush} (gs);
  \draw[arr] (sn2) -- (in1);
  \draw[arr] (in1.north) -- (obj.south);
  \draw[arr] (obj.east) -- (ss.west);
  \draw[arr] (gs.east)  -- ++(0.2,0) |- (qn.west);
  \draw[arr] (ss.south) -- (qn.north);

  \node[font=\tiny, gray, below=0pt of gs] {searchable $<$1\,s};
  \node[font=\tiny, gray, below=0pt of ss] {ANN-indexed};
\end{tikzpicture}
\caption{Log-structured write path (Milvus-style~\cite{bib:milvus}).
  All writes are durably committed to the WAL.  Growing Segments provide
  immediate brute-force searchability; Index Nodes asynchronously build
  ANN-optimized Sealed Segments persisted to object storage.}
\label{fig:logpath}
\end{figure}

A log-structured write path separates write durability from index construction.
All writes are committed first to a durable, ordered Write-Ahead Log (WAL);
index structures are derived asynchronously.  Figure~\ref{fig:logpath}
illustrates the architecture.

\textbf{Milvus}~\cite{bib:milvus} uses a WAL (Apache Pulsar or Kafka) as the
sole source of truth.  Each log entry carries a Timestamp Oracle (TSO)
timestamp for snapshot-isolation reads.  \emph{Growing Segments} are in-memory,
brute-force searchable immediately after WAL acknowledgment.  When a Growing
Segment reaches a size threshold, dedicated \emph{Index Nodes} build HNSW or
IVF vector indexes and flush to object storage as immutable \emph{Sealed
Segments}.  The four worker types---Streaming Nodes (ingest), Query Nodes
(search), Index Nodes (ANN construction), and Data Nodes (flush)---scale
independently.

\textbf{Havenask / HA3}~\cite{bib:havenask} (Alibaba's production search
engine for Taobao/Tmall) uses a \emph{Build Service} operating in three modes:
(i)~\emph{full build} produces a complete index periodically (daily) and
delivers it to Searchers via HDFS; (ii)~\emph{incremental build} processes
message-queue updates, delivering deltas every 30--60\,min; (iii)~\emph{real-time
mode} runs Build Service as an in-process library inside each Searcher,
building index structures directly into Searcher memory for second-level
freshness.  Single-machine benchmarks report $4\times$ higher QPS and
$4\times$ lower latency than Elasticsearch on comparable datasets.

\emph{Trade-offs.}  The write path (WAL write) is immediately durable, while
query visibility depends on the construction path (sub-second for Growing
Segments; minutes to hours for offline builds).  Operational complexity
increases (WAL management, TSO service), but the query path is completely
isolated from index construction.

\subsection{In-Place Partial Updates}

Segment creation and merging are necessary for full-text fields.  However,
scalar values (prices, timestamps, stock levels, category IDs) used only for
filtering and scoring do not need the segment model.  For these fields, a
much cheaper update path is possible.

\textbf{Vespa's Proton engine}~\cite{bib:vespa_attr} formalizes this as a
first-class design.  \emph{Attribute fields} (scalar/vector values) are stored
as forward-index arrays in RAM, updated in-place in $O(1)$ time with no segment
creation, no merge overhead, and immediate query visibility.  String fields use
an Enum Store (32-bit integer references) to deduplicate values.  When the
\texttt{fast-search} option is enabled, an in-memory B-tree provides $O(\log n)$
range filtering with copy-on-write semantics for concurrent read-write safety.

Attribute updates are first appended to a sequential Transaction Log Service
(TLS) for durability, then applied synchronously to memory, making changes
visible to queries before the TLS \texttt{fsync}.  HNSW vector graphs for
tensor attributes are also updated incrementally in memory, requiring no full
graph rebuild~\cite{bib:vespa_hnsw}.

\emph{Trade-offs.}  Zero merge latency and instant query visibility for
non-text fields; limited to fields that do not require linguistic analysis.
Text index fields still use the standard segment model.

\subsection{Offline Build with Hot-Swap}

When update latency requirements are measured in hours rather than seconds,
a \emph{periodic offline rebuild} approach may be preferred.  The index is
built entirely offline on separate hardware from a data snapshot; the serving
cluster is updated atomically by swapping in the new index.

The serving cluster operates exclusively from a static, fully merged index with
no background merge activity, yielding maximally predictable latency and optimal
query throughput.  \emph{Hot-swap}~\cite{bib:wandering} maintains two copies of
the index simultaneously during transitions (double memory), migrating queries
atomically when the new copy is fully loaded.

\emph{Trade-offs.}  Best-possible query latency stability, at the cost of update
freshness (limited by build cycle duration) and double-memory overhead during
transitions.  Appropriate for large, slow-changing catalogs.

\section{Comparative Analysis}

\subsection{System Summary}

Table~\ref{tab:comparison} compares representative systems across five
dimensions.  No architecture dominates all dimensions simultaneously:
\begin{itemize}[leftmargin=*]
  \item \textbf{Node-level separation} trades propagation latency for query
        latency stability.
  \item \textbf{Compute-storage separation} trades cold-start latency for
        perfect horizontal scalability and zero write-read interference.
  \item \textbf{Full in-memory} trades DRAM cost for lowest and most
        predictable query latency.
  \item \textbf{Log-structured paths} trade index construction latency for
        write isolation; Growing Segments recover sub-second visibility.
  \item \textbf{In-place attributes} trade data-model constraints for
        zero-latency updates on non-text fields.
\end{itemize}

\begin{table*}[t]
\centering
\caption{Comparison of write-read decoupling approaches across representative
  systems.}
\label{tab:comparison}
\small
\renewcommand{\arraystretch}{1.15}
\begin{tabular}{@{}lllcll@{}}
\toprule
\thead{System / Architecture} &
\thead{Update\\Freshness} &
\thead{Query\\Latency} &
\thead{Write-Read\\Isolation} &
\thead{Serving\\Memory} &
\thead{Relative\\Cost} \\
\midrule
ES (shared-nothing)            & $\sim$1\,s           & Variable (spikes)    & None      & Moderate     & High (WAF 10--30$\times$) \\
LinkedIn Galene (node sep.)    & Minutes              & Stable, low          & High      & Disk + cache & Moderate \\
Uber Sia (node sep.)           & Minutes              & Stable, low          & High      & Disk + cache & Moderate \\
Quickwit (compute-storage)     & Minutes--hours       & Stable, moderate     & Complete  & Cache only   & Very low \\
ES Stateless / AI Lake         & Seconds--minutes     & Stable, low          & Complete  & Cache only   & Low \\
Alibaba OpenStore              & Seconds              & Stable, low          & Complete  & Cache only   & Low \\
Algolia (full in-memory)       & $\sim$1\,s           & Very low, stable     & High      & Very high    & High \\
Typesense (full in-memory)     & Instant              & Very low, stable     & High      & Very high    & High \\
Ximalaya ad index              & Seconds              & Very low ($<$5\,ms)  & High      & High         & High \\
Milvus (log backbone)          & Sub-second (growing) & Low, stable          & Complete  & Mixed        & Moderate \\
Vespa (attr.\ in-place + idx)  & Instant (attr.)      & Very low, stable     & High      & Moderate     & Moderate \\
Havenask HA3                   & Seconds (RT mode)    & Very low (4$\times$) & High      & High         & High \\
\textbf{ScaleSearch}           & Sub-minute           & Extremely low        & Complete  & Very high    & Moderate--High \\
\bottomrule
\end{tabular}
\end{table*}

\subsection{Selection Criteria}

The appropriate architecture depends on workload characteristics:
\begin{itemize}[leftmargin=*]
  \item \textbf{High-concurrency large-scale search / ad bidding ($<$10\,ms P99, minute-level freshness):}
        Full in-memory or compute-storage + in-memory synthesis (ScaleSearch).
  \item \textbf{Log / observability analytics at massive scale:}
        Compute-storage separation (Quickwit, ES frozen tier); immutable
        append-only data suits the immutable split model.
  \item \textbf{Mixed workloads with frequent attribute updates:}
        Vespa's attribute-plus-index separation.
  \item \textbf{Large catalog with periodic update cycles:}
        Offline build with hot-swap.
  \item \textbf{Real-time vector search:}
        Log-structured path (Milvus); Growing Segments provide immediate
        brute-force visibility; Sealed Segments provide efficient ANN.
\end{itemize}

\section{The ScaleSearch Design: A Synthesis}

ScaleSearch is an in-house search engine targeting high-concurrency large-scale
product retrieval at billion-document scale with requirements of P99 query
latency under 50\,ms, sub-minute update freshness, and per-query recall in the
tens of thousands.

\begin{figure}[t]
\centering
\begin{tikzpicture}[node distance=0.25cm]
  \node[sbox=2.0cm, fill=gray!15, minimum height=0.55cm] (client)
    at (1.8, 3.8) {Client (Queries / Writes)};

  \node[sbox=1.5cm, fill=yellow!25, minimum height=0.6cm] (master)
    at (1.8, 2.85) {Master Node\\(etcd-backed)};

  \node[sbox=1.3cm, fill=orange!25] (wn1) at (0, 1.6) {Write Node};
  \node[sbox=1.3cm, fill=orange!25] (wn2) at (0, 0.7) {Write Node};

  \node[sbox=1.3cm, fill=blue!20, minimum height=1.3cm] (s3)
    at (1.8, 1.15) {Object\\Storage\\(S3)};

  \node[sbox=1.35cm, fill=green!20] (rn1) at (3.65, 1.6) {Search Node\\(full RAM)};
  \node[sbox=1.35cm, fill=green!20] (rn2) at (3.65, 0.7) {Search Node\\(full RAM)};

  \draw[arr]  (client) -- (master);
  \draw[darr, gray] (master.west)  -- ++(-.35,0) |- (wn1.north);
  \draw[darr, gray] (master.west)  -- ++(-.35,0) |- (wn2.north);
  \draw[darr, gray] (master.east)  -- ++( .35,0) |- (rn1.north);
  \draw[darr, gray] (master.east)  -- ++( .35,0) |- (rn2.north);

  \draw[arr, orange!70]
    (wn1.east) -- node[font=\tiny, above, sloped]{upload} (s3.west);
  \draw[arr, orange!70] (wn2.east) -- (s3.west);

  \draw[arr, green!60]
    (s3.east) -- node[font=\tiny, above, sloped]{load / poll} (rn1.west);
  \draw[arr, green!60] (s3.east) -- (rn2.west);

  \node[font=\tiny, gray, below=1pt of wn2] {Lucene segs → upload};
  \node[font=\tiny, gray, below=1pt of rn2] {in-memory inverted idx};
\end{tikzpicture}
\caption{ScaleSearch architecture.  Write Nodes build Lucene segments and
  upload them to object storage; Search Nodes load all segments into full
  in-memory indexes and poll for new segments.  The two paths share no
  hardware resources and scale independently via the Master (etcd-backed).}
\label{fig:scalesearch}
\end{figure}
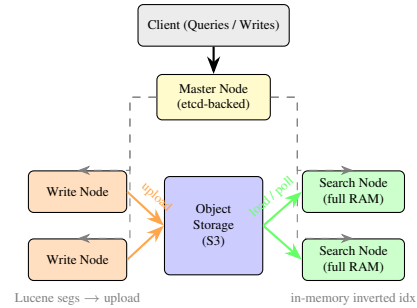

The central design insight is that fields within a document have
\emph{fundamentally different write patterns}.  A product's title and
description change rarely and require linguistic analysis; its price, stock
level, and promotion tags change continuously and require only scalar storage.
No prior system exposes this distinction as a first-class routing mechanism at
field granularity.  ScaleSearch introduces \emph{per-field update routing}: each
field owns a dedicated Kafka topic and is assigned one of two update paths
matched to its semantics.  This single mechanism subsumes and unifies
node-level separation (Section~4.1), compute-storage separation
(Section~4.2), full in-memory indexing (Section~4.3), and in-place partial
updates (Section~4.5) within one coherent architecture.
Figure~\ref{fig:scalesearch} shows the overall topology.

\subsection{Architecture}

\textbf{Write path.}  A dedicated \emph{Write Node} per shard receives document
updates and runs a custom \emph{segment writer}---a Lucene-inspired component
that produces immutable, self-describing segment files (term dictionary,
posting lists, doc values, stored fields) without coupling to the Lucene
library.  The Write Node performs no large-scale merges; after sealing a
segment locally, it uploads the segment files to shared object storage
asynchronously.

\textbf{Read path.}  \emph{Search Nodes} are stateless with respect to
durability.  On startup, they download all segment files for their assigned
shards from object storage and use a custom \emph{segment reader} to construct
a full in-memory inverted index---loading term dictionaries, posting lists, and
forward arrays directly into native heap structures with no \texttt{mmap} or
OS page-cache dependency.  Subsequently, they poll object storage for new
segment files; when new files appear, they are downloaded and merged into the
in-memory index via a pure in-memory merge (no disk I/O, no page-cache
disruption).

\textbf{Cluster management.}  etcd stores shard metadata and drives leader
election.  A Master node coordinates shard assignment, node health monitoring,
and index lifecycle management.

\subsection{Per-Field Update Routing: Key Innovation}
\label{sec:perfield}

Real-world applications impose \emph{heterogeneous freshness requirements}
across fields of the same document.  A product's inventory and promotion tags
must reflect changes within seconds to avoid overselling or expired promotions;
its title and description, which change rarely and require tokenization, can
tolerate minute-level staleness.  No existing system exposes this distinction as
a first-class mechanism: Elasticsearch processes all fields together in each
segment flush; Vespa distinguishes attribute from text fields but shares the
same serving node and TLS for both.

ScaleSearch introduces two complementary abstractions to address this:

\textbf{Column families.}  Fields are grouped into \emph{column families}
($\mathit{CF}$) according to their freshness SLA.  All fields in the same
column family share a Kafka consumer group, a common update cadence, and the
same update path.  A typical deployment defines a small number of column
families---e.g., \textsf{CF-realtime} (sub-second freshness) for scalar signals
and \textsf{CF-text} (minute-level freshness) for full-text fields.  This
grouping is the operationally natural unit: engineers set freshness SLAs at the
field-family level rather than per field, and the system enforces them uniformly
within each family.

\textbf{Per-field Kafka topics.}  Within a column family, each field owns a
dedicated Kafka topic.  This allows independent replay, rate control, and lag
monitoring per field, while sharing the broader update pipeline of the column
family.

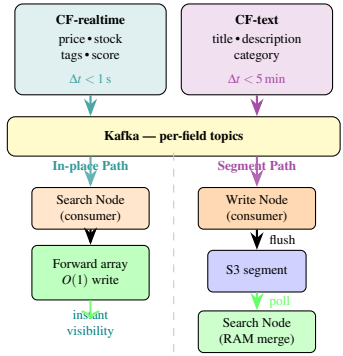
\begin{figure}[t]
\centering
\begin{tikzpicture}
  \node[draw, fill=teal!12, rounded corners=3pt,
        minimum width=2.0cm, minimum height=1.2cm,
        align=center, font=\tiny, inner sep=3pt] (cfrt) at (0.9, 4.15)
    {\bfseries CF-realtime\\[1pt]price\,\textbullet\,stock\\
     tags\,\textbullet\,score\\[3pt]\textcolor{teal!75}{$\Delta t <$\,1\,s}};

  \node[draw, fill=violet!12, rounded corners=3pt,
        minimum width=2.0cm, minimum height=1.2cm,
        align=center, font=\tiny, inner sep=3pt] (cftxt) at (3.1, 4.15)
    {\bfseries CF-text\\[1pt]title\,\textbullet\,description\\
     category\\[3pt]\textcolor{violet!70}{$\Delta t <$\,5\,min}};

  \node[draw, fill=yellow!25, rounded corners=3pt,
        minimum width=4.4cm, minimum height=0.52cm,
        font=\tiny\bfseries] (kafka) at (2.0, 3.0)
    {Kafka --- per-field topics};

  \draw[arr, teal!70]   (cfrt.south)  -- (cfrt.south  |- kafka.north);
  \draw[arr, violet!70] (cftxt.south) -- (cftxt.south |- kafka.north);

  \node[font=\tiny\bfseries, teal!80]   at (0.9, 2.6) {In-place Path};
  \node[sbox=1.55cm, fill=orange!20] (scons) at (0.9, 2.05)
    {Search Node\\(consumer)};
  \node[sbox=1.55cm, fill=green!25, minimum height=0.72cm] (rarr) at (0.9, 1.2)
    {Forward array\\$O(1)$ write};
  \node[font=\tiny, teal, align=center] at (0.9, 0.52)
    {instant\\visibility};

  \draw[arr, teal!60]   (kafka.south -| scons) -- (scons.north);
  \draw[arr] (scons) -- (rarr);
  \draw[->, green!55, thick] (rarr.south) -- ++(0,-0.25);

  \node[font=\tiny\bfseries, violet!70] at (3.1, 2.6) {Segment Path};
  \node[sbox=1.55cm, fill=orange!30] (wnode)  at (3.1, 2.05)
    {Write Node\\(consumer)};
  \node[sbox=1.3cm,  fill=blue!18]   (s3seg)  at (3.1, 1.22)
    {S3 segment};
  \node[sbox=1.55cm, fill=green!20]  (sload)  at (3.1, 0.42)
    {Search Node\\(RAM merge)};

  \draw[arr, violet!60] (kafka.south -| wnode) -- (wnode.north);
  \draw[arr] (wnode) -- node[font=\tiny, right=1pt]{flush} (s3seg);
  \draw[arr, green!60] (s3seg) -- node[font=\tiny, right=1pt]{poll} (sload);

  \draw[dashed, gray!45] (2.0, 0.15) -- (2.0, 2.82);
\end{tikzpicture}
\caption{Per-field update routing in ScaleSearch.  Fields are grouped into
  \emph{column families} by freshness SLA.  \textsf{CF-realtime} fields
  (left) are consumed directly by Search Nodes and written into a forward array
  in $O(1)$ with instant visibility.  \textsf{CF-text} fields (right) are
  consumed by a Write Node, flushed as Lucene segments to object storage, then
  merged into the Search Node's in-memory inverted index.}
\label{fig:perfield}
\end{figure}

Figure~\ref{fig:perfield} illustrates how the two column families map to two
distinct update paths:

\textbf{In-place path (\textsf{CF-realtime}).}  Fields such as
\texttt{price}, \texttt{stock\_level}, and \texttt{rank\_score} are stored as
forward-index arrays in RAM, one slot per local document ID.  A Search Node
consumes the field's Kafka topic directly and writes the new value to the
array slot in $O(1)$ time, with no segment creation and no merge.  The update
is immediately visible to concurrent queries.  String fields use an Enum Store
for value deduplication.

\textbf{Segment path (\textsf{CF-text}).}  Fields such as \texttt{title},
\texttt{description}, and \texttt{category\_text} require tokenization,
FST-based term dictionaries, and positional posting lists.  Their updates are
consumed by a Write Node, which runs the custom segment writer and flushes the
resulting segment files to object storage.  Search Nodes poll object storage,
download new segments, and use the segment reader to merge them into the
in-memory inverted index---a pure RAM operation bounded in cost by the new
segment size.

The routing path for each field is declared at index-creation time in the field
mapping.  In practice, a product document typically has $O(10)$ text fields on
the segment path and $O(100)$ scalar fields on the in-place path.  The vast
majority of production update traffic---continuous signals such as click-through
rates, inventory counts, and promotion weights---therefore never reaches the
Write Node or object storage at all.

\textbf{Comparison with prior work.}  Vespa's Proton engine (Section~4.5)
distinguishes attribute from text fields, but attribute updates still flow
through the Transaction Log Service on disk before being applied to memory, and
both paths share the same serving node.  ScaleSearch removes this residual
coupling: the in-place path is a direct memory write on the Search Node, with
no disk serialization; the segment path is physically isolated on a separate
Write Node.  The column-family abstraction further makes freshness SLAs an
explicit, operationally enforced contract rather than an emergent property of
refresh-interval tuning.

\subsection{How It Solves Write-Read Contention}

The design achieves isolation through four mechanisms:

\begin{enumerate}
  \item \textbf{Structural isolation.}  Search Nodes never run the segment
        writer.  All write-path work (tokenization, segment construction,
        upload) occurs exclusively on Write Nodes.
  \item \textbf{Storage decoupling.}  Object storage is the only coupling point
        between write and read paths.  Search Nodes can be added or replaced
        without data migration; scaling read capacity has no effect on write
        load.
  \item \textbf{Memory-based merge on the read side.}  When a Search Node
        merges a new segment into its in-memory index, the operation is a pure
        in-memory computation: no disk writes, no \texttt{fsync}, no
        page-cache disruption.  Its cost is bounded by the \emph{new} segment
        size, not the total index size.
  \item \textbf{Per-field routing eliminates unnecessary write overhead.}
        The majority of production update traffic consists of scalar field
        changes (prices, inventory, scores).  By routing these directly to
        in-place RAM updates on Search Nodes---bypassing the Write Node, S3
        upload, and segment merge entirely---ScaleSearch avoids generating any
        write amplification for the most frequent update type.  Only structural
        text changes, which are far less frequent, traverse the segment path.
\end{enumerate}

\subsection{Memory-Aware Sharding}

A full in-memory index requires that each shard fit within a single node's
DRAM.  ScaleSearch adopts HBase-style range splits~\cite{bib:bigtable}: shards
are defined by key ranges and split when their memory footprint exceeds a
configurable threshold.  This ensures every shard remains within one node's
capacity while allowing horizontal scale-out by adding Search Nodes and
splitting shards---without global rehashing or cross-cluster data rebalancing.

\subsection{Design Rationale}

The core engineering judgment is that DRAM cost to hold the full index is
justified by the reduction in query latency and operational complexity compared
to disk-based alternatives.  Cloud economics support this: the cost of
serving-tier DRAM (per query-latency-ms saved) is often lower than the cost of
managing disk I/O contention, replication write amplification, and
merge-related latency incidents in a conventional Elasticsearch cluster.

Update freshness is bounded by the upload pipeline (seconds) plus the Search
Node polling interval (configurable, typically 10--30\,s), achieving
sub-minute freshness that meets the target SLA.

\section{Challenges and Future Directions}

\subsection{Hybrid Full-Text and Vector Search}

Modern search systems increasingly require simultaneous lexical matching
(BM25, inverted index) and semantic matching (approximate nearest neighbor,
HNSW/IVF).  Vector index updates are structurally different from and more
expensive than inverted index updates, creating new write pressure patterns.

\emph{Filtered ANN} search---retrieving the top-$k$ nearest neighbors subject
to a predicate filter---has no universally satisfactory solution~\cite{bib:vdbsurvey}.
Lucene 9.3 introduced HNSW pre-filtering (LUCENE-10606); Vespa has supported
filtered HNSW for several years.  Multi-stage ranking pipelines~\cite{bib:lin2021}---a
fast first-stage retriever (BM25 or dense bi-encoder) followed by a BERT
cross-encoder reranker---decouple the memory-bound, latency-critical retrieval
stage from the compute-intensive reranking stage, enabling independent scaling.

\subsection{Serverless and Elastic Scaling}

Elastic Cloud Serverless (GA late 2024) realizes the compute-storage separation
pattern at the product level.  Vexless~\cite{bib:vexless} (SIGMOD 2024)
explores serverless vector search using ephemeral cloud functions as stateless
query workers, achieving zero cost at rest.  The same model applies to
full-text search: a ``search function'' loads only relevant splits per query,
yielding true pay-per-query semantics.

\subsection{Hardware Trends}

\textbf{CXL memory pooling} allows multiple servers to share a common DRAM pool
via a high-bandwidth, low-latency interconnect.  For full in-memory search, CXL
could allow a cluster to collectively hold an index larger than any single
node's DRAM.  Sherman~\cite{bib:sherman} demonstrates the feasibility of
distributed B$^+$-tree indexes over disaggregated RDMA memory; analogous
results for inverted indexes would directly enable the ScaleSearch model to
scale beyond single-node memory limits.

\subsection{AI-Integrated Search}

Retrieval-Augmented Generation (RAG)~\cite{bib:lin2021} uses a classical
in-memory index as the retrieval backbone; real-time document ingestion
generates standard write pressure.  At the frontier, Differentiable Search
Indices (DSI)~\cite{bib:dsi} encode the document collection into model
weights, potentially eliminating explicit index structures.  However, model
weight updates require expensive fine-tuning, and there is no equivalent of
the NRT update model for neural index structures.  GIR faces its own version
of the write-read tension: how to incorporate new documents without expensive
retraining.

\section{Conclusion}

Write-read contention is a fundamental challenge in large-scale search engine
design, rooted in the write amplification and background merge activity
inherent in the LSM-based Lucene segment model and amplified by the
shared-nothing replication model of first-generation distributed architectures.
This survey identifies and analyzes five principal decoupling patterns: node-level
read-write separation, compute-storage separation, full in-memory indexing,
log-structured write paths, and in-place partial updates.  Each makes a
different trade-off in the freshness-latency-cost space.

The ScaleSearch architecture synthesizes compute-storage separation with full
in-memory indexing, achieving complete structural isolation between write and
read paths, horizontal scalability without data migration, and sub-millisecond
query latency for high-concurrency large-scale consumer search.

The core principle---\emph{the write path and the read path should be isolated
at every layer of the system stack}---will remain the guiding design principle
for high-performance search engines as vector search integration, serverless
deployments, and AI-augmented retrieval create new forms of write pressure in
the coming decade.


\end{document}